\documentclass[usenatbib]{mnras}

\usepackage[utf8]{inputenc}
\usepackage{graphicx}
\usepackage{subfigure}
\usepackage{amsmath}
\usepackage{siunitx}
\usepackage{placeins}
\usepackage{comment,xcolor}
\usepackage{appendix}
\usepackage{caption}
\usepackage{gensymb}
\usepackage{float}
\usepackage{ulem}
\usepackage{xcolor}
\usepackage{bm}

\definecolor{dg}{rgb}{0.0, 0.6, 0.1}

\newcommand{\bfm}[1]{\mbox{\boldmath$ #1 $}}
\graphicspath{{Pictures/}}
\title[MHD Fermi bubbles]{The role of magnetic fields in shaping $\gamma$-ray emission from the Fermi bubbles}

\author[O.~Tourmente et al.]{
Olivier Tourmente$^{1}$\thanks{E-mail: olivier.asin@desy.de},
Donna Rodgers-Lee$^{2}$ and
Andrew M.~Taylor$^{1}$
\\
$^{1}$Deutsches Elektronen-Synchrotron DESY, Zeuthen, Germany\\
$^{2}$Astronomy \& Astrophysics Section, School of Cosmic Physics, Dublin Institute for Advanced Studies, 31 Fitzwilliam Place, \\Dublin D02 XF86, Ireland\\ }
\date{Accepted XXX. Received YYY; in original form ZZZ}
\pubyear{2025}

\begin{document}
\maketitle

\begin{abstract}
Despite their discovery fifteen years ago, the nature and origin of the Fermi bubbles remain unclear. We here investigate the effect a magnetic field can have on a subsonic breeze outflow emanating from the Galactic centre region. The presence of this magnetic field allows anisotropic diffusion of cosmic rays within the outflow, shaping the resultant cosmic ray distribution obtained out at large distances within the Galactic halo. We show that our magnetohydrodynamic Galactic breeze model, in combination with an opening angle for the injection of cosmic rays, leads to $\gamma$-ray emission from the Fermi bubble region with relatively sharp edges. 
\end{abstract}


\section{Introduction}

In 2010, the discovery of large-scale diffuse Galactic emission in the form of two lobes extending above and below the Galactic centre was reported \citep{Su_2010, Dobler_2010}, utilising data from the $\gamma-$ray satellite Fermi-LAT. These features indicate the presence of a Galactic-scale outflow, extending up to $\sim$10~kpc above the Galactic plane, which are now known as the Fermi bubbles. Some key features of this emission are a near-constant surface brightness intensity, relatively sharp edges of the emission, and an approximately power-law spectral energy distribution described by $dN/dE \propto E^{-1.9 \pm 0.2}$ \citep{Fermi_2014}. The width of the base of the Fermi bubbles appears to coincide with the microwave haze \citep{Finkbeiner_2004, Planck_2013}. Moreover, recent observations made in the X-ray band have revealed even larger bubbles, called the eROSITA bubbles, which enshroud the Fermi bubble structures \citep{eROSITA_2020}, and have been suggested to also originate from a Galactocentric outflow \citep{Zhang:2024mcd}.

Observations of UV absorption lines from cold clouds in the Fermi bubbles have been used to trace their line-of-sight velocity and indicate their direction of propagation. These observations have revealed an underlying velocity profile for these clouds \citep{Fox_2015, Bordoloi_2017, Karim_2018, Lockman_2020, Ashley_2020, Cashman_2021, Sofue_2022}. The profile appears to reach a maximum velocity of $\sim$300~\rm{km s$^{-1}$} at a Galactic latitude of $\sim 10^{\circ}$, before decelerating continuously with increasing Galactic latitude.

To date, no consensus regarding the physical mechanism responsible for producing the Fermi bubbles has yet been reached \citep[for a review, see][] {Sarkar_2024}. Several candidate mechanisms, however, have been proposed. Hadronic $\gamma$-ray emission wind models \citep{Crocker_2011, Mou_2014, Mou_2015, Sarkar_2017} and leptonic jet models \citep{Guo_2012, Yang_2012, Yang_2013, Yang_2017, Guo_2017}, have primarily focused on the existence of a supersonic Galactic outflow out at Fermi bubble size scales \citep[though see][as exceptions to this]{Mertsch_2011,Crocker_2015,Shimoda:2024ogb}. However, the velocity profile of the gas for a supersonic outflow scenario appears to be incompatible with the slowly decreasing outflow velocities inferred using UV absorption line techniques. 

In addition to the velocity profile, the gas density profile provides information on the nature of the outflow. A supersonic outflow implies a rapid decrease of gas density with increasing distance in the Galactic halo \citep{Sarkar_2017, Sarkar_2021} due to the dominance of ram pressure over thermal pressure in the outflow. However, recent observations of O~\textsc{\small VII} absorption measurements and ram pressure stripping observations \citep{Martynenko_2022}, Sunyaev-Zeldovich measurements \citep{Bregman_2022}, and thermal X-ray emission \citep{Zhang:2024zwu}, have provided new insights on the Galactic halo density distribution. These results all find consistency with expectations for the halo being supported by thermal pressure in hydrostatic equilibrium \citep{Faerman_2017,Prochaska_2019}.

A third model, consistent with the density distribution of the Galactic halo gas being in approximate hydrostatic equilibrium, considers a subsonic outflow model, called a Galactic breeze \citep{Taylor_2017}. Building on this, the propagation of a thermally driven outflow reproducing a subsonic solution \citep{Bondi_1952, Parker_1958} was simulated using a hydrodynamic model \citep{Tourmente_2023}. This work showed that a Galactic breeze solution is capable of producing a Galactic bubble with a decelerating velocity profile, giving rise to $\gamma$-ray emission which broadly matches the Fermi-LAT observations \citep{Fermi_2014}. However, in this model the $\gamma$-ray emission was produced in a broader region than the Fermi bubble $\gamma$-ray emission. 

Magnetic fields in the outflow may alter several aspects of a hydrodynamic outflow description. Indeed, observations of Galactic synchrotron emission indicate that magnetic fields permeate the Fermi bubbles, with a strength of 6-12 $\mu$G for the case of volume-filled lobes and 13-15 $\mu$G for the case of shell-dominated emission \citep{Carretti_2013}. These findings are consistent with a recent study that looked at Planck data polarised synchrotron emission maps and inferred the presence of a $\sim$ 7 $\mu$G within a spatially extended region of $\sim$ 6~kpc above the Galactic centre \citep{Shaw_2022}. Here, we extend our previous hydrodynamic model to a magnetohydrodynamic (MHD) model and investigate the effects introduced by the presence of magnetic fields.

This paper is structured as follows: in Section \ref{sec:Galactic_outflow} the MHD equations are introduced, as well as the effect of a magnetic field on a thermally driven outflow and the numerical setup. In Section \ref{sec:CR_transport} the cosmic ray transport equation used is described. The results are presented in Section \ref{Results}, followed by a discussion in Section \ref{Discussion}. Our conclusions are presented in Section \ref{Conclusion}.

\section{Galactic outflow model}\label{sec:Galactic_outflow}

An MHD simulation is performed for the propagation of the subsonic outflow into the hot halo region with the PLUTO code \citep{Pluto}. The ideal MHD module solves the mass continuity (Eq.~\ref{Mass_conservation}), momentum continuity (Eq.~\ref{Momentum_conservation}), and induction equation (Eq.~\ref{Faraday_law}) as a function of time:
\begin{align}
    &\frac{\partial \rho}{\partial t} + \bm{\nabla}\cdot \left(\rho \textbf{\textit{v}}\right) = S_{\rho},\label{Mass_conservation}\\
    &\frac{\partial}{\partial t}\left(\rho \textbf{\textit{v}}\right) + \bm{\nabla}\cdot \left(\rho \textbf{\textit{v}} \textbf{\textit{v}} + P\textbf{\textit{I}}\right) - \nabla \cdot \left(\frac{\textbf{B}\textbf{B}}{4\pi} - \frac{B^2\textbf{I}}{8\pi}\right) = -\rho \bm{\nabla} \Phi,\label{Momentum_conservation}\\
    &\frac{\partial \textbf{\textit{B}}}{\partial t} - \bm{\nabla}\times\left(\bm{v} \times \bm{B}\right) = 0. \label{Faraday_law}
\end{align}
where $\rho$ is the mass density, $\textbf{\textit{v}}$ is the velocity vector, $S_{\rho}$ is a source term representing the mass injection rate (see Section~\ref{Boundary}), $P$ is the thermal pressure, $\textbf{\textit{I}}$ is the unitary tensor, $\textbf{\textit{B}}$ is the magnetic field vector, and $c$ is the speed of light. The effect of the total Galactic gravitational potential, $\Phi$ (see Section~\ref{Galactic_potential}), is represented by the term on the right-hand side of Eq.~(\ref{Momentum_conservation}). On the left-hand side of Eq.~(\ref{Momentum_conservation}), the second term corresponds to the ram and thermal pressure. The third term corresponds to the magnetic tension and the magnetic pressure. A second-order linear spatial reconstruction with a van Leer limiter has been used in combination with a second-order Runge-Kutta scheme to advance the equations in time. An HLL Riemann solver \citep{HLL} is used for computing the intercell fluxes. The divergence-free condition for the magnetic field is maintained using the divergence cleaning scheme \citep{Div_clean}. We assume, as a first approximation, that the gas in the Galactic halo has an isothermal temperature distribution (see Section~\ref{Galactic_potential}), and we therefore omit the energy continuity equation.

\subsection{Setup and Initial Conditions}\label{Initial}

The computational domain is discretised with a 2.5D grid using spherical coordinates ($r,\theta$). The number of grid points in the $r$-direction is $N_\mathrm{r}=256$ bins, and in the $\theta$-direction, $N_\mathrm{\theta}= 92$ bins. The inner radius is $r_{0}=0.3$~kpc, and the outer radius is $r_\mathrm{max}=300$~kpc. For the $\theta$-direction, the inner angle is $\theta_{0}=0$ and the outer angle is $\theta_{\mathrm{max}}=\pi/2$. 

In Section~\ref{Galactic_potential} the gravitational potential model and the density distribution are presented. Section~\ref{Boundary} presents the setup chosen for the initial values of the magnetic field, velocity and mass density at the inner radial boundary.\newline

\begin{figure}
    \centering
    \includegraphics[width=1.0\linewidth]{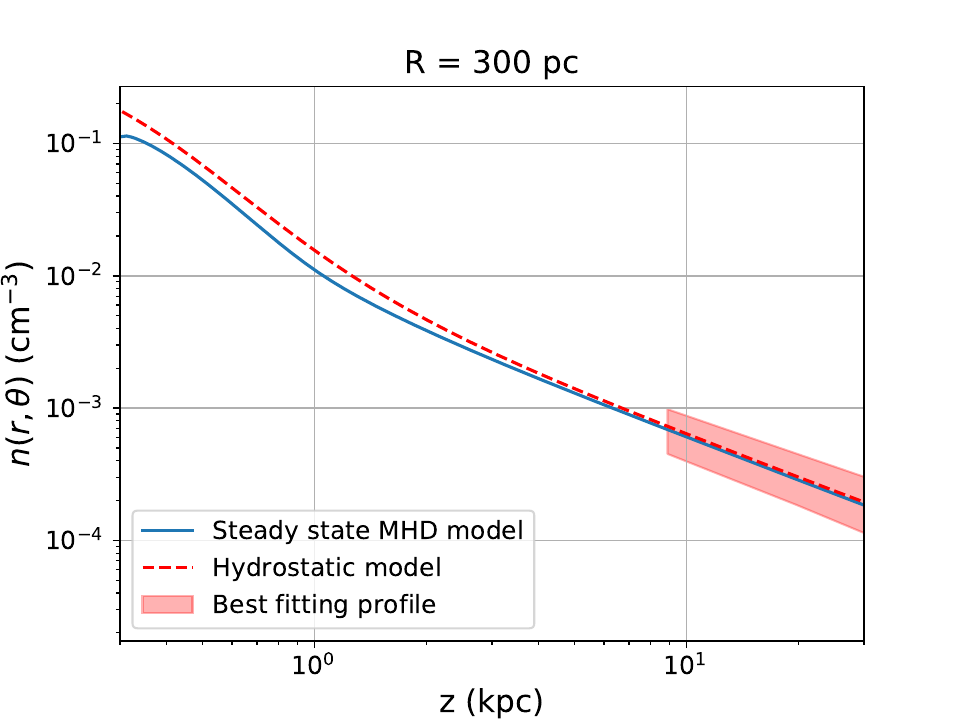}
    \caption{Gas density distribution of the hot Galactic halo as a function of Galactic height for R = 0.3~kpc. The dashed red line represents the hydrostatic density distribution for the Galactic halo (Eq.~\ref{eq:rho}). The solid blue line represents the steady-state density distribution for the numerical simulation. The red shaded region represents a fitting range provided by observations of the O~\textsc{\small VII} spectrum and ram pressure stripping \citep{Martynenko_2022}.}
    \label{fig:Density}
\end{figure}

\subsubsection{Galactic gravitational potential and density distribution}\label{Galactic_potential}

Similar to \citet{Tourmente_2023}, the total Galactic gravitational potential represents the sum of three components: a bulge ($\Phi_{\rm b}$), disc ($\Phi_{\rm d}$) and dark matter halo ($\Phi_{\rm DM})$ such that
\begin{equation*}
    \Phi = \Phi_{\rm b} + \Phi_{\rm d} + \Phi_{\rm DM}.
\end{equation*}
For each component, we adopt the same definitions as in \citet[Section 2.1]{Tourmente_2023}. The gravitational potential has been normalised by using the fitting profile provided for the Milky Way by \citet{Watkins_2019}, based on Gaia observations \citep{Gaia}. Fig. 1 from \citet{Tourmente_2023} shows $\Phi$ as a function of Galactic radius and a comparison with the fitting range deduced from the Gaia observations.

The initial number density distribution representing the hot Galactic halo gas is expressed as a hydrostatic density distribution given by 
\begin{equation}
    n_{\mathrm{gas}} = n_{10} \exp{\left( -\frac{\Phi - \Phi_{10}}{c^2_s}\right)},
    \label{eq:rho}
\end{equation}

where $n_{\rm gas}=\rho/m_{p}$, and $n_{10} = 3\times 10^{-3}$~cm$^{-3}$ is the number density at a distance, $r_{10}=10$~kpc from the Galactic centre. $\Phi_{10}$ represents the value of the gravitational potential at $r=10$~kpc. This value has been chosen to match with observations of the O~\textsc{\small VII} spectrum and ram pressure stripping \citep{Martynenko_2022}, as shown in Fig.~\ref{fig:Density}. As the Galactic halo is approximated with an isothermal temperature distribution, the thermal velocity, $c_s$, is constant and is chosen to be $250$~km~s$^{-1}$, corresponding to an energy of $ kT \sim400$~eV. The limitations of this model assumption are further discussed in Section~\ref{Discussion}. A density distribution for the Galactic disc has not been included in our initial condition setup, but its gravitational influence is present through $\Phi_{\rm d}$. The scale height for the Galactic disc is 300~pc \citep[][Table 1]{Tourmente_2023}. This is also the position of our inner radial boundary of the simulations, which means that the effect of neglecting the Galactic disc density should be negligible. The effect of neglecting the disc gas density in our initial condition setup is discussed in Section~\ref{CR_injection}.

Fig. \ref{fig:Density} shows both the initial density distribution of the gas (corresponding to hydrostatic equilibrium) for the numerical simulation and the late time steady-state distribution. As thermal pressure in the subsonic outflow dominates over ram pressure, there is very little evolution of the density distribution with time. The steady-state density distribution (solid blue line) is similar to the initial density distribution. As the hot Galactic halo is considered to be isothermal, the thermal pressure can be expressed as $P = \rho c_s^{2}$.

\subsubsection{Boundary conditions}\label{Boundary}

At the beginning of the simulation, the velocity and magnetic field strength are set to zero throughout the computational domain, except at the inner radial boundary. At the inner radial boundary, $r_{0}$, an initial value has been set up for both radial and azimuthal components for velocity ($v_{r_{0}}, v_{\phi_{0}}$) and magnetic field ($B_{r_0}, B_{\phi_0}$) as well as the mass injection rate ($\dot{M}$). The outer radial boundary condition has been set up with an outflow condition. Both the inner and outer $\theta$ boundaries have been set up with a reflective condition.

The initial radial velocity corresponds to the value calculated analytically to reach a maximum Mach number of M=0.85 at a Galactic radius of 1~kpc. This is $v_{r_{0}}=211$~km~s$^{-1}$, the same as that adopted for our earlier hydrodynamic model \citep{Tourmente_2023}. For the initial azimuthal velocity, $\sim$10\% of the rotational velocity of the Galactic disc has been chosen, corresponding to $v_{\phi_{0}} = 20$~km~$\rm{s}^{-1}$ \citep{Honma_2012, Honma_2015}. The initial value for $\dot{M}$ depends on the initial value of both the velocity and gas density. The chosen value for $v_{r_{0}}$ corresponds to $\dot{M}=0.75~M_{\odot}$~yr$^{-1}$. This value lies in the upper range of the estimated star formation rate, $0.03-1 M_{\odot}$~yr$^{-1}$, at the Galactic centre \citep{Henshaw_2023}. In combination with the initial value for the velocity, the injected kinetic energy luminosity can be calculated, for which one obtains $L_{E_{k}}\approx 10^{40}$~erg~s$^{-1}$.

For our simulations the initial (inner radial) magnetic field values of $B_{r_{0}}$ and $B_{\phi_{0}}$ have been chosen so as to not significantly disrupt the velocity profile. For the radial magnetic field an initial value of $B_{r_{0}}=20~\mu$G has been chosen. The initial value for the azimuthal magnetic field, $B_{\phi_{0}} = 2~\mu$G, has been chosen. Further consideration of the role of magnetic fields in Eq. \ref{Momentum_conservation} is left to appendix~\ref{App:Bfield}. In particular, it is highlighted that $B_{\phi}$ can potentially influence the velocity profile. This happens through the magnetic tension term when the magnetic field lines, initially pointing to the $r$-direction, bend over and point in the $\phi$-direction. 

\subsection{MHD simulation results}
\label{Pluto_Numerical_setup}

\begin{figure*}%
\centering
\subfigure[]{%
   \includegraphics[width=0.45\textwidth]{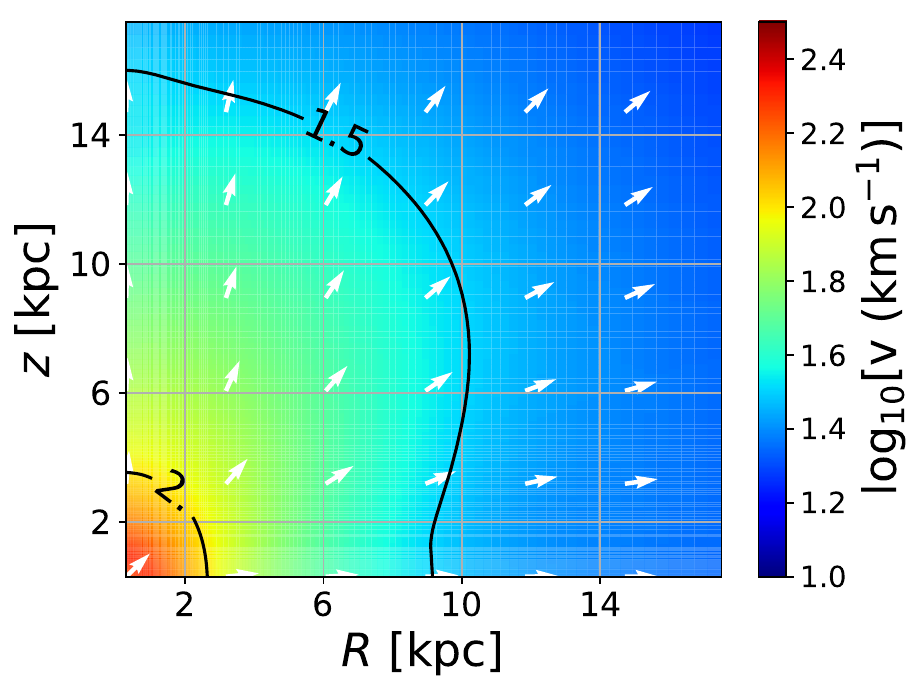}
    \label{fig:_a}}%   
\centering
\subfigure[]{%
   \includegraphics[width=0.45\textwidth]{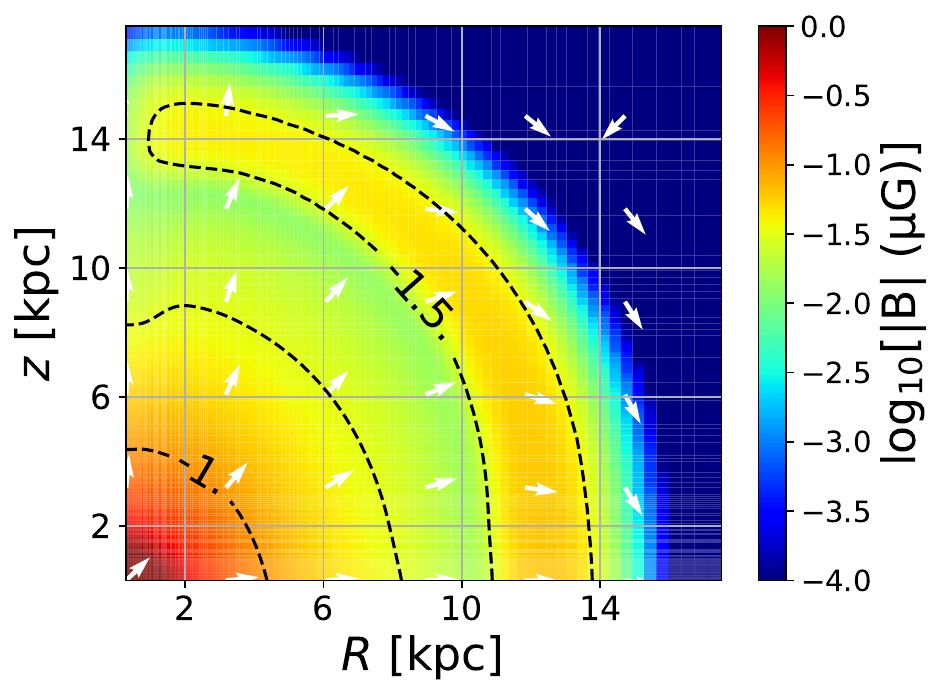}
    \label{fig:_b}}%  
\\
~
\subfigure[]{%
   \includegraphics[width=0.45\textwidth]{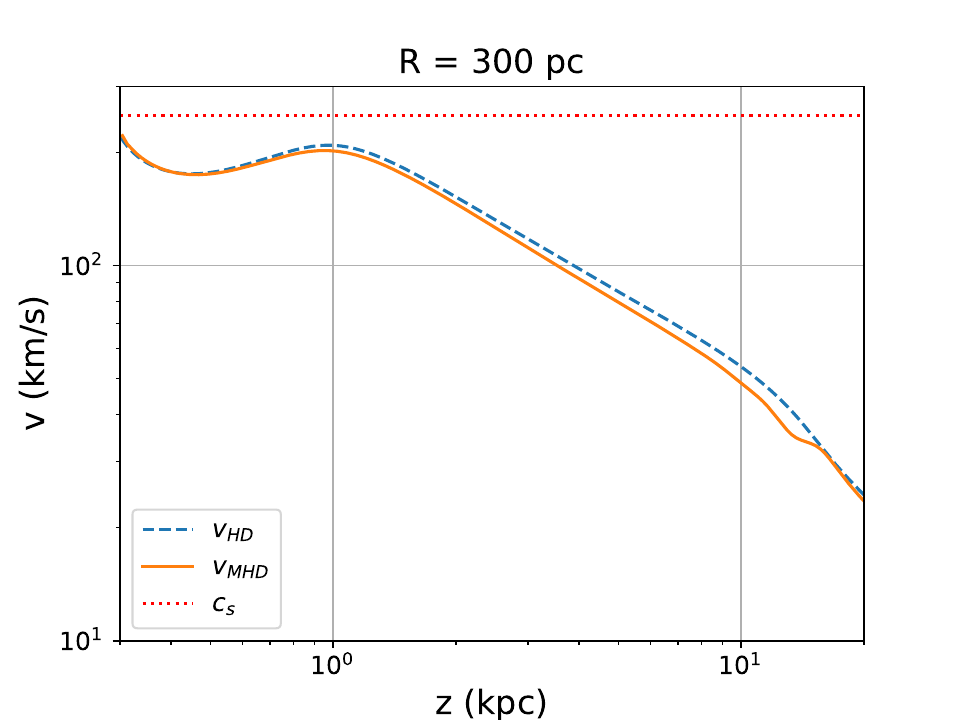}
    \label{fig:_c}}%   
\centering
\subfigure[]{%
   \includegraphics[width=0.45\textwidth]{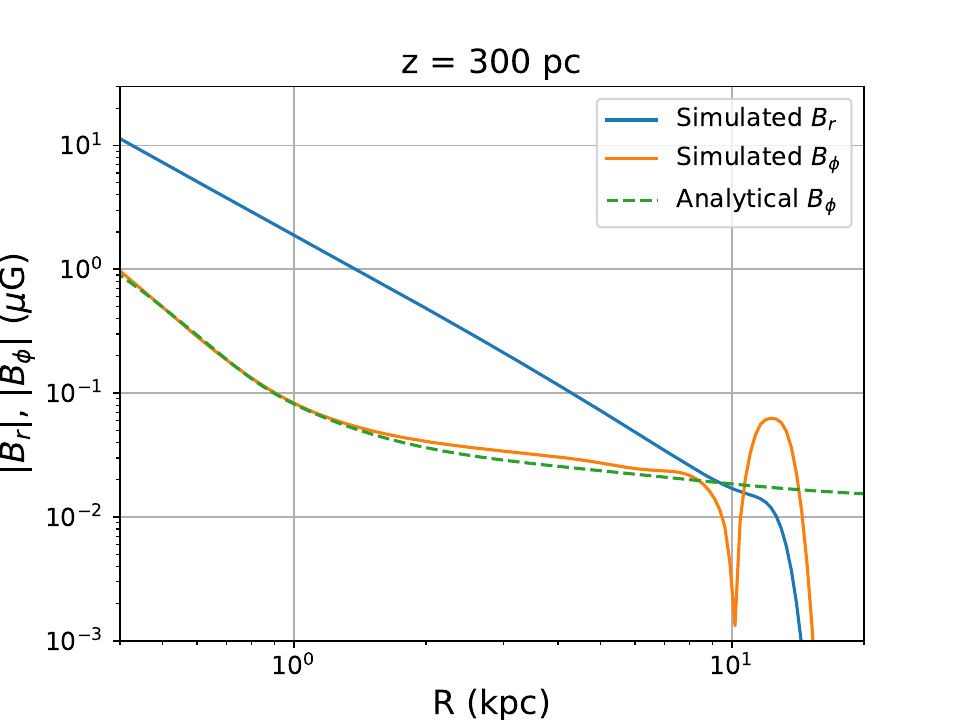}
    \label{fig:_d}}%

\caption{Spatial distribution of the steady-state (a) velocity and (b) magnetic field strength, both for a 2D map with cylindrical coordinates. The white arrows represent the direction of the respective vector field. The continuous colour bar indicates the logarithm in base 10 for each quantity. For Fig.~\ref{fig:_a} the black contour lines represent the velocity distribution for two specific values of 100~km~s$^{-1}$ and 30~km~s$^{-1}$. For Fig.~\ref{fig:_b} the black contour lines represent magnetic field strengths of 0.1~$\mu$G and 0.03~$\mu$G. Fig.~\ref{fig:_c} shows the distribution of the velocity as a function of Galactic height, $z$, for $R=300$~pc. The solid orange line represents the velocity profile from the MHD simulation, and the dashed blue line represents the velocity profile for the hydrodynamic solution. The dotted red line represents the thermal velocity.
Fig.~\ref{fig:_d} shows the distribution of both magnetic field components, $B_r$ and $B_{\phi}$, along the cylindrical radius, $R$, for $z=300$~pc. The solid blue and orange lines represent the distribution of $B_r$ and $B_{\phi}$. For comparison, the dashed green line represents the analytic solution for the distribution of $B_{\phi}$.}
\label{fig:MHD_simu}
\end{figure*}

The two panels in Fig.~\ref{fig:MHD_simu} show the velocity and magnetic field distributions in steady-state out to $r=10$~kpc, obtained from the numerical outflow simulation with the setup presented in Section~\ref{Initial}. The colour bar in Figs.~\ref{fig:_a}-\ref{fig:_b} denotes the logarithm of the velocity and magnetic field strengths, respectively. Fig.~\ref{fig:_a} presents the 2D velocity distribution as a function of the cylindrical coordinates ($R$, $z$). Two contour lines (solid black lines) have been drawn, one for a velocity of 100~km~s$^{-1}$ and one for 30~km~s$^{-1}$. 
%These MHD results produce a velocity profile very similar to our previously considered hydrodynamical model results \citep[section 4.1]{Tourmente_2023}. 
The velocity reaches a maximum of $\sim 210$~km~s$^{-1}$ at a Galactic radius of 1~kpc and slowly decelerates continuously beyond this point. 

Fig.~\ref{fig:_b} presents the 2D magnetic field distribution as a function of position in cylindrical coordinates ($R$, $z$). The two contour lines shown in the figure (dashed black lines) correspond to 0.1~$\mu$G and 0.03~$\mu$G. For the first 10~kpc from the Galactic centre, the magnetic field strength decreases continuously with increasing distance.

Fig.~\ref{fig:_c} and Fig.~\ref{fig:_d} show the velocity and magnetic field, respectively. For both Fig.~\ref{fig:_c} and Fig.~\ref{fig:_d}, the components $v_{\theta}$ and $B_{\theta}$ are negligible in comparison of $v_r$, $v_{\phi}$, $B_r$, and $B_{\phi}$, respectively, and have therefore not been included. As the Fermi bubbles extend mainly along the $z$-direction, Fig.~\ref{fig:_c} emphasises the velocity distribution along the Galactic height, $z$, for $R=300$~pc. Fig.~\ref{fig:_d} emphasises the $B_r$ and $B_{\phi}$ distribution along the cylindrical radius for $z=300$~pc. For Fig.~\ref{fig:_c}, the velocity profile produced by the MHD model is shown by the solid orange line, with the dashed blue line showing the corresponding result obtained for the hydrodynamic model \citep{Tourmente_2023}. As appreciated from the similarity of these results, the MHD result describes a subsonic velocity profile broadly similar to that obtained for the hydrodynamical model.

For Fig.~\ref{fig:_d} the distribution of $B_r$ is shown by the solid blue line and by a solid orange line for $B_{\phi}$. Following the divergence-free condition on the magnetic field, $B_r$ decreases as $\propto r^{-2}$. By considering the induction equation, Eq.~(\ref{Faraday_law}), in steady-state it is possible to obtain an analytic expression for $B_{\phi}$ that can be compared to the $B_{\phi}$ distribution produced by the numerical simulation. The different steps that lead to this expression are left in appendix \ref{App:Bphi}. It is given by
%]
\begin{equation}\label{Frozen-in}
    B_{\phi}(r) = \left[B_{\phi_{0}} + B_{r_{0}}\frac{v_{\phi_{0}}}{v_{r_{0}}}\left(\frac{r_0^2}{r^2} - 1\right)\right]\left(\frac{r_0}{r}\right)^{1-\alpha},
\end{equation}
where $\alpha$ is a power-law exponent that depends on the slope of $v_r(r)$. The magnetic field is anchored to the outflow that will be influenced in turn by the magnetic tension or magnetic pressure. 

As seen in Fig.~\ref{fig:_d}, out to a radius of 10~kpc from the Galactic centre, the distribution of $B_{\phi}$ obtained with the MHD simulation matches well with the analytical distribution (dashed green line) obtained with Eq.~(\ref{Frozen-in}). A shell structure is formed at a distance of 12-14~kpc from the Galactic centre, out at the radius where $B_{\phi}$ becomes larger than $B_r$. At radii larger than this distance, the magnetic field is azimuthal in direction, and the velocity flow is no longer along the magnetic field lines. This misalignment of the velocity flow and magnetic field results in the primarily the advection, and a lesser extent the compression, of the azimuthal magnetic field, resulting in the formation of the azimuthal magnetic field shell. Appendix \ref{App_Induction} presents the influence of the compression and advection terms of the induction equation that lead to the formation of the shell.
However, it should be noted that for our simulations this shell is not in steady-state at the end of the simulation time and is still evolving.

\section{Cosmic ray transport model}\label{sec:CR_transport}

The cosmic ray (CR) transport equation that we adopt to model their propagation through the Galactic outflow is expressed as
\begin{equation}\label{CRs_transport}
    \frac{\partial f}{\partial t} = \bfm{\nabla}\cdot\left(\bfm{D}\cdot\nabla f - \bfm{v} f\right) + \frac{1}{p^2}\frac{\partial }{\partial p}\left[\left(\bfm{\nabla}\cdot \bfm{v}\right)\frac{p^3}{3}f\right] - \frac{f}{\tau_{\rm loss}} + \frac{Q}{p^2},
\end{equation}
where $f$ is the CR space phase density, given by $f=\frac{dN}{d^3x d^3p}$, where $dN$ is the number of particles within a volume of size $d^{3}xd^{3}p$. The outflow velocity, $\textbf{\textit{v}}$, is obtained from the MHD simulation (see Section \ref{sec:Galactic_outflow}), $p$ is the CR momentum, $\tau_{\rm loss}$ is the CR energy loss time scale through inelastic proton-proton ($pp$) collisions (see Section \ref{Loss_time_scale}) and $Q$ represents the number of CRs injected per unit of time and per unit of volume (see Section \ref{CR_injection}). $\bfm{D}$ is the anisotropic diffusion tensor which is discussed in Section \ref{Diffusion}.

The first term on the right-hand side of Eq.~(\ref{CRs_transport}) represents the divergence of the total CR spatial current. The total CR spatial current itself has two contributions, the diffusive current, which depends on the strength and level of turbulence of the magnetic field (see Section \ref{Diffusion}) and the advective current. The second term on the right-hand side represents CR momentum advection that describes the evolution of the CR momentum during their transport within the fluid. The third term on the right-hand side describes CR energy losses through inelastic collisions with ambient gas \citep{Gabici_2007}. 

The temporal evolution of the CR distribution function $f$, obtained using the (steady-state) MHD results for the velocity profile for the advection term and the magnetic field distribution for the diffusion term, was obtained using a spatial differencing scheme \citep{Rodgers_Lee_2017, Rodgers_Lee_2020}. The computational grid adopted for the CR transport used a cylindrical ($R$, $z$) grid with azimuthal symmetry, with 60 logarithmically spaced spatial grid points in $R$ and $z$. Both the $R$ and $z$ axes ranged from 0.01~kpc to 17.5~kpc. 

Since the MHD simulation grid range started from radius $r_{0}$ (300~pc), an interpolation of the velocity and magnetic field distribution was initially made onto the cylindrical CR transport grid. 
For radii on the CR grid that are smaller than those on the MHD grid (i.e., $<r_{0}$), the velocity and magnetic fields obtained from the MHD grid were extrapolated onto the CR grid.

The CR momentum grid ranges from 10~GeV/c to 30~GeV/c with 5 logarithmically spaced momentum bins. The simulations have been run for a simulation time of 300~Myr, i.e., until a steady-state is reached out to a radial distance of $r\approx 10$~kpc. This distance corresponds approximately to the height reached by the Fermi bubbles. The inner spatial boundaries, $R_{\mathrm{in}}$ and $z_{\mathrm{in}}$, have both been set up with a reflective boundary condition. For the outer spatial boundaries, $R_{\mathrm{out}}$ and $z_{\mathrm{out}}$, both have been set up with an outflow boundary condition. The inner and outer momentum boundaries have been set up as outflow \citep{Rodgers_Lee_2020}.

Simulations of CR transport have been performed for only positive $z$. For Figs.~\ref{CR_Density_map} and \ref{fig:skymap}, showing CR density distribution and $\gamma$-ray emission for both the north and south Galactic hemispheres in Section \ref{Results} (see also appendix~\ref{App:30_40} and \ref{Appendix_iso}) the results have been reflected in the $z=0$ axis since we assume the system is symmetric about the $z$-axis, an equivalent set of results would also be expected for the negative $z$ region.

\begin{figure*}%
\centering
\subfigure[]{%
   \includegraphics[width=0.45\textwidth]{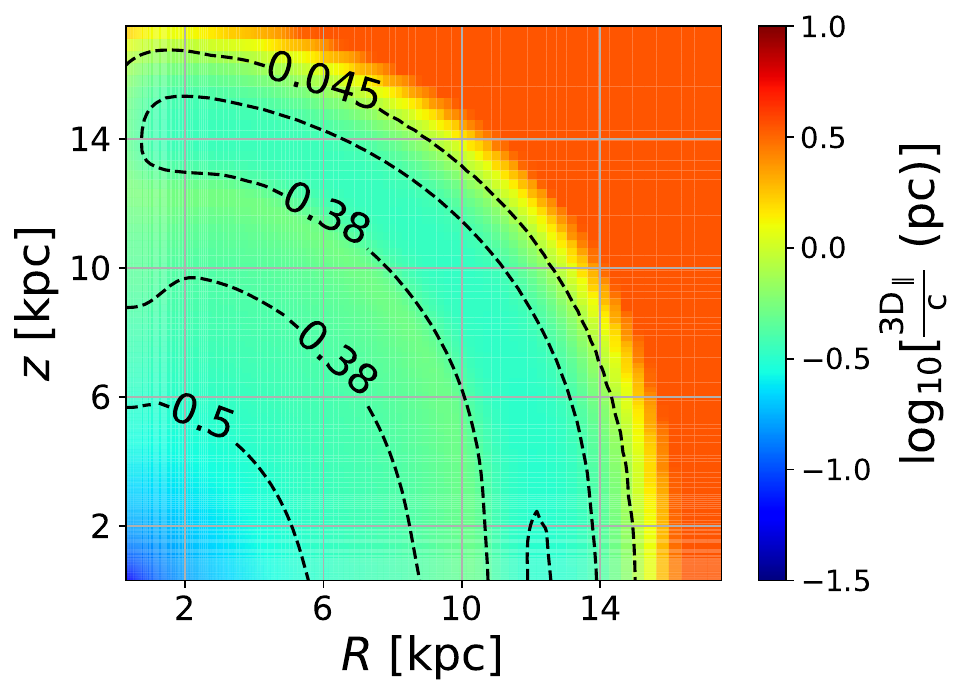}
    \label{fig:_a_diff}}%   
\centering
\subfigure[]{%
   \includegraphics[width=0.45\textwidth]{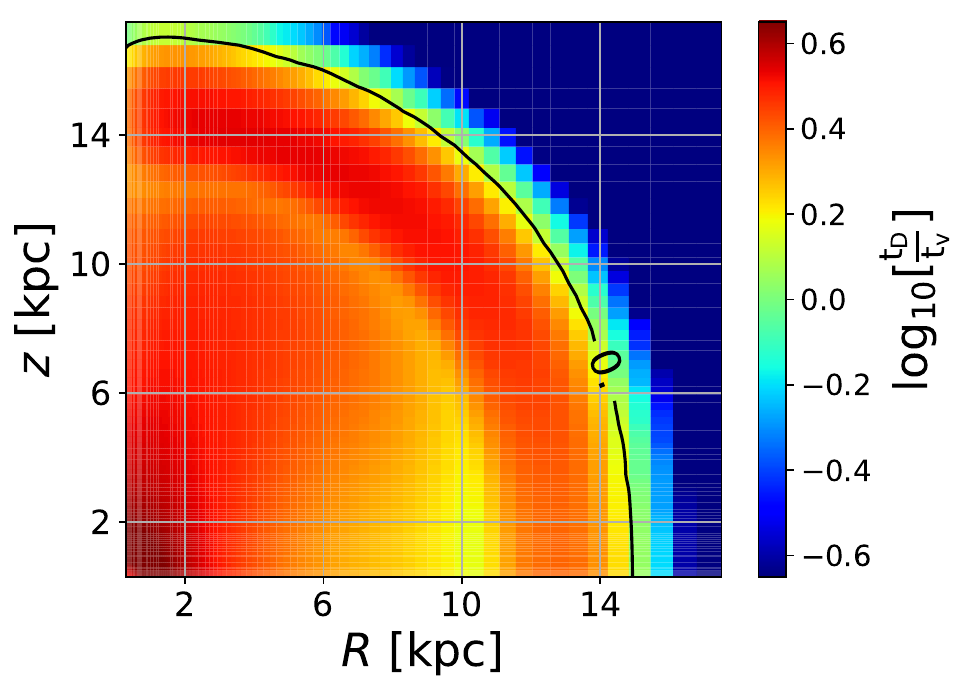}
    \label{fig:_b_diff}}%  

\caption{Spatial distribution of (a) the diffusion scattering length and (b) the ratio of the diffusion time to the advection time. The continuous colour bar indicates the logarithm in base 10. (a) The dashed contour lines represent the diffusion scattering length, $3D_{\parallel}(30~{\rm GeV})/c$, for three different values, 0.3~pc , 0.42~pc and 0.9~pc. (b) The solid contour line represents the position where the ratio of the diffusion time is equal to the advection time.}
\label{fig:}
\end{figure*}

\subsection{CR source}\label{CR_injection}
 
In order to match with observations that suggest the central outflow responsible for the formation of the Fermi bubbles is collimated, we inject CRs in our simulations at the inner radial boundary within a half opening angle, $\theta_{1/2} = 20^{\circ}$, measured from the $z$-axis. This value was motivated by observations of the Fermi bubbles and their central chimneys, i.e., a structure that appears to funnel hot gas, dust and CRs into the Galactic halo, present at the Galactic centre. For the Fermi bubbles, $\theta_{1/2} \approx 18^{\circ}-30^{\circ}$, given that the size of the Fermi bubbles is estimated to be $\sim 8~-~10$~kpc in height and $\sim 3-4$~kpc for the half-width \citep{Carretti_2013, Fermi_2014}. An X-ray chimney at the base of the Fermi bubbles has been observed with a half-width of $\sim 50$~pc and a height of $\sim 150$~pc \citep{Ponti_2019}. Similarly, a chimney was observed in radio emission with a half-width of $\sim 70$~pc and a height of $\sim 215$~pc \citep{Heywood_2019}. Assuming a conic angle for the distribution, those two structures have $\theta_{1/2}\approx 20^{\circ}$. The interaction of the outflowing gas with the disc density material, although not included in our simulations (see Section~\ref{Galactic_potential}), may be the agent responsible for the collimation of the outflow. However, we leave an investigation into this to future work.

CRs are injected through a source term, $Q$, onto the computational grid, with a momentum distribution that assumes a power-law shape, expressed as,
\begin{equation}\label{injection_Q}
Q = \frac{d\dot{N}}{dp} = \frac{\dot{N}_{\rm min}}{p_{\rm min}}\left(\frac{p}{p_{\rm min}}\right)^{-\eta},%\exp{\left(-\frac{p}{p_{\mathrm{max}}}\right)},
\end{equation}
where $p_{\rm min} = 1.876$~GeV/$c$, $\eta = 2$  \citep[representative of diffusive shock acceleration][]{Bell_1978}. $\dot{N}_{\rm min}$ is a constant of normalisation representing the number of particles injected per second with momentum $p_{\rm min}$. The value of $\dot{N}_{\rm min}$ sets the CR luminosity value. For this work, $L_{\mathrm{CR}} = 1.3\times 10^{40}$~erg~s$^{-1}$ has been chosen after comparing the produced $\gamma$-ray emission with observations (see Section \ref{Gamma_ray_emission_map}). Given that it is assumed here that CRs originate from the Galactic centre, CRs are injected into the region $10~\mathrm{pc} < r < 315~\mathrm{pc}$. 

\subsection{Diffusion and advection}\label{Diffusion}

The distance that a CR can travel, on average, until its momentum vector is significantly changed (by $\sim$ 1~rad. in angle) by scattering off a magnetic field is determined by the diffusion tensor, $\bfm{D}$, and is expressed as,
\begin{equation}
\bfm{D} = D_{\perp}\left(\bfm{I} - \bfm{\hat{b}}\bfm{\hat{b}}\right) + D_{\parallel}\bfm{\hat{b}}\bfm{\hat{b}}. \label{eq:D}
\end{equation}
%
%In \citet{Tourmente_2023}, an isotropic and homogenous diffusion coefficient was adopted. 
 The parallel diffusion coefficient, $D_{\parallel}$, represents the diffusion along the magnetic field unit vector direction, $\bfm{\hat{b}}\equiv \bfm{B}/|\bfm{B}|$. The perpendicular diffusion coefficient, $D_{\perp}$, represents the diffusion in the plane perpendicular to the magnetic field direction. Eq.\,\ref{eq:D} is discussed in further detail in Appendix \ref{Appendix_B}. Given that the MHD simulation provides the distribution of the magnetic field, we explore the effect of adopting an anisotropic and inhomogeneous diffusion coefficient.

From quasi-linear theory \citep{DeMarco_2007}, the parallel diffusion scattering length, $3D_{\parallel}/c$ is given by \citep{Jokipii_1966, Schlickeiser_1989},

\begin{equation}\label{Dc_inhomo}
\frac{D_{\parallel}}{c} = \frac{B^2_c}{\delta B^2}\left(\frac{r_L}{\lambda_{\mathrm{max}}}\right)^{1-\zeta}r_{L},
\end{equation}
where $\zeta = 5/3$ represents a Kolmogorov-type turbulence spectrum. $\lambda_{\mathrm{max}}$ represents the longest wavelength mode in the turbulence, for which we adopt $\lambda_{\rm max}=$1.5~pc. $r_{L}$ is the CR Larmor radius. $\delta B ^{2}/B_c^{2}$ describes the ratio of the energy density in the turbulent magnetic field to that in the coherent magnetic field. The ratio of the energy density is set to unity, i.e., $B_c^{2} = \delta B^{2}$. 

Similarly, from quasi-linear theory \citep{DeMarco_2007}, the  perpendicular diffusion coefficient is given by,
\begin{equation}
\frac{D_{\perp}}{c}= \frac{\delta B^2}{B_c^2}\left(\frac{r_L}{\lambda_{\rm max}}\right)^{\zeta -1}r_{L}.
\end{equation}
 However, the validity of quasi-linear theory for describing the perpendicular diffusion coefficient is particularly questionable  \citep{Giacalone_1999,Casse_2001,Cohet_2016}, for which field line wandering is suggested to also play a key role in particle transport \citep{Mertsch_2020}. We therefore instead appeal to results from approaches which attempt to go beyond this description as a motivation for the value of $D_{\perp}/D_{\parallel}$ we adopt.

Several works have provided a ratio between the perpendicular and parallel diffusion scattering lengths $\sim 10^{-2}-10^{-1}$, either directly through numerical simulations \citep{Casse_2001,Hussein_2015, Cohet_2016}, or by attempts to derive new analytical forms for the scattering rates \citep{Shalchi_2010,Gammon_2017}. Here, as a first step, we arbitrarily select a ratio of $D_{\perp} = 0.1D_{\parallel}$, which sits within this motivated range.

The magnetic field from the MHD simulations (section \ref{Pluto_Numerical_setup}) was used to define the diffusion tensor orientation in each simulation cell. Since the turbulent magnetic field is random, and coherent on considerably smaller scales than the simulation cells, its contribution to $\bfm{\hat{b}}$ was neglected.

Fig.~\ref{fig:_a_diff} shows the 2D distribution of the parallel diffusion scattering length defined by Eq.~(\ref{Dc_inhomo}) as a function of $R$ and $z$. The contour lines (dashed black line) have been drawn for a diffusion scattering length of $3D_{\parallel}(30~{\rm GeV})/c=0.3$~pc, $3D_{\parallel}(30~{\rm GeV})/c=0.42$~pc and $3D_{\parallel}(30~{\rm GeV})/c=0.9$~pc. Since the parallel diffusion scattering length depends on the magnetic field distribution, its profile can be easily understood with Figs.~\ref{fig:_b} and \ref{fig:_d}. As the radial magnetic field component $B_r$ dominates over $B_{\phi}$ centrally in the simulation and decreases in strength as $\propto r^{-2}$, the parallel diffusion scattering length increases as the radius increases. Beyond a radius of $\sim$ 14~kpc, $B_{\phi}$ forms a thin shell which alters how the parallel diffusion scattering length changes with distance. Beyond a radius of $r>$17~kpc, a minimal (floor) value for the magnetic field has been set to $2.6\times 10^{-5}$~$\mu$G, corresponding to $3D_{\parallel}(30~\mathrm{GeV})/c = 3$~pc.

Within each cell the dominant transport process is governed by the competition between the advection and diffusion rates in that cell. Fig.~\ref{fig:_b_diff} shows the spatial distribution of the ratio between the diffusion time, $t_{\mathrm{D}} = (\Delta r)^2/(D_{\parallel})=(\Delta \ln r)^2 r^{2}/(D_{\parallel})$, and the advection time, $t_{\mathrm{v}} = \Delta r/v= (\Delta \ln r)r/v $, where $\Delta \ln r$ is constant due to the logarithmic spatial grid adopted. A contour line (solid black line) has been drawn on this figure indicating when the ratio of these timescales is unity, i.e., where the diffusion time equals the advection time. Due to the small value of the total magnetic field strength (see Fig.~\ref{fig:MHD_simu}), the advection time is shorter than the diffusion time in the region formed by the outflow (between 0.3~kpc and $\sim$16~kpc). This suggests that advection is the dominant CR transport mechanism. However, the magnetic field distribution still imprints itself on the results through its impact on anisotropic CR diffusion. 

\subsection{Energy loss time scale and $\gamma$-ray emission}\label{Loss_time_scale}

Inelastic $pp$ collisions of CR protons in the outflow with the thermal Galactic halo gas lead to the creation of charged and neutral pions. The neutral pion subsequently decays into two photons \citep{2020PTEP.2020h3C01P}. 

The CR energy loss time scale, $\tau_{\mathrm{loss}}$, is controlled by the density distribution of the Galactic halo gas and is given by \citep{Gabici_2007}
\begin{equation}\label{subsec:Energy_loss}
    \tau_{\mathrm{loss}} = \frac{1}{c \kappa \sigma_\mathrm{pp} n_{\rm gas}},
\end{equation}
where $\kappa \approx 0.5$ represents the inelasticity of the CR in the $pp$ collision, $\sigma_\mathrm{pp} \approx 4\times 10^{-26}$~cm$^{2}$ is the total inelastic cross-section for $pp$ collisions, and $n_{\rm gas}$ is the target gas number density (section \ref{Pluto_Numerical_setup}). It should be noted that the value for $\tau_{\mathrm{loss}}$ is larger than the simulation time for most CR on the simulation grid due the low density of the hot Galactic halo gas. The value for the loss time, however, imprints itself on the brightness of subsequent $\gamma$-ray emission produced (see Eq.~\ref{Energy_flux}).

Through inelastic CR collisions with ambient gas, approximately $\sim 10~\%$ of the relativistic proton's energy goes into each produced $\gamma$-ray photon \citep{Kelner:2006tc}. As the CRs considered here have an energy range of 10 to 30~GeV, photons with an energy range of $\sim 1-3$~GeV are therefore produced. The sum of all CR energy losses within the background gas gives the total non-thermal $\gamma$-ray luminosity, $L_{\gamma}$ produced, which in turn is set by the CR injection rate onto the grid, represented by $L_{\mathrm{CR}}$. 

For an observer at Earth, the sum of all the $\gamma$-ray emission produced along a solid angle of angular size $\Delta \Omega$, results in the total $\gamma$-ray brightness, $E_{\gamma}F_{\gamma}$. This $\gamma$-ray brightness is expressed by 
\begin{equation}\label{Energy_flux}
    E_{\gamma}F_{\gamma} = \frac{1}{3\Delta \Omega \tau_{\mathrm{loss}}} \int \frac{e_{\mathrm{CR}}}{r^2_{\mathrm{obs}}}dV = \frac{1}{12\pi \tau_{\mathrm{loss}}} \int e_{\mathrm{CR}} dr_{\mathrm{obs}},
\end{equation}
where $r_{\mathrm{obs}}$ is the distance between the source of emission and the observer. $dV$ represents the conical volume element within the solid angle. The factor of $\frac{1}{3}$ accounts for the fraction of neutral pions produced in each inelastic CR interaction. For the simulations, the position of the observer from the Galactic centre has been fixed at 8~kpc \citep{Gillessen_2017}. The corresponding CR energy density, $e_{\mathrm{CR}}$, can be obtained by integrating over the particle momentum distribution, 
\begin{equation}
    e_{\mathrm{CR}} = \int^{\mathrm{p_{max}}}_{\mathrm{p_{min}}} 4\pi f p^3 dp.
\end{equation}

\section{Results}\label{Results}

\begin{figure}
    \includegraphics[width=1.\linewidth]{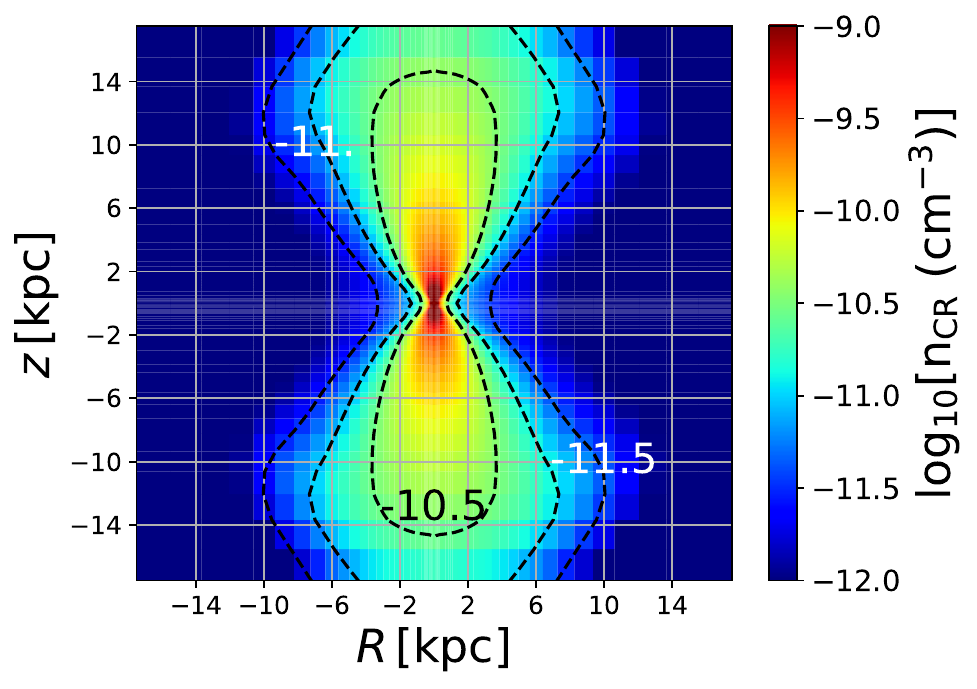}
    \caption{Spatial CR density distribution in the Galactic halo for an energy range of $E_{\mathrm{CR}} = 10~-~30$~GeV with the colour bar showing the logarithm in base 10.  The black dashed black lines are contours representing different density distributions, i.e., $n_{\mathrm{CR}} = 0.3, 1$ and $3\times 10^{-11}$~cm$^{-3}$, respectively.}
    \label{fig:Density_distrib}
\end{figure}
The CR propagation results obtained using the MHD Galactic outflow simulations are described in Section \ref{CR_Density_map}. The resultant $\gamma$-ray maps, produced through $pp$ collisions between CRs and the Galactic halo gas, are presented in Section \ref{Gamma_ray_emission_map}.   

\subsection{CR density map}\label{CR_Density_map}

The CR transport simulations were run until the CR distribution reached a steady-state up to a Galactic height of $\sim 10$~kpc, corresponding to $\sim 300$~Myr. The transport of CRs has been simulated based on the velocity and magnetic field distribution obtained from the Galactic breeze model (see Section \ref{sec:Galactic_outflow}). Fig.~\ref{fig:Density_distrib} shows a cross-sectional slice of the resultant late-time CR density distribution, plotted in cylindrical coordinates. The CR density distribution ranges from $10^{-12}$~cm$^{-3}$ (dark blue) to $10^{-9}$~cm$^{-3}$ (dark red). Three contour lines are shown in Fig.\,\ref{fig:Density_distrib} (dashed black lines), representing three different CR density values, $n_{\mathrm{CR}} = 0.3, 1$, and $3\times 10^{-11}$~cm$^{-3}$, respectively. 

\begin{figure*}
    \center
    \includegraphics[width=0.74\linewidth]{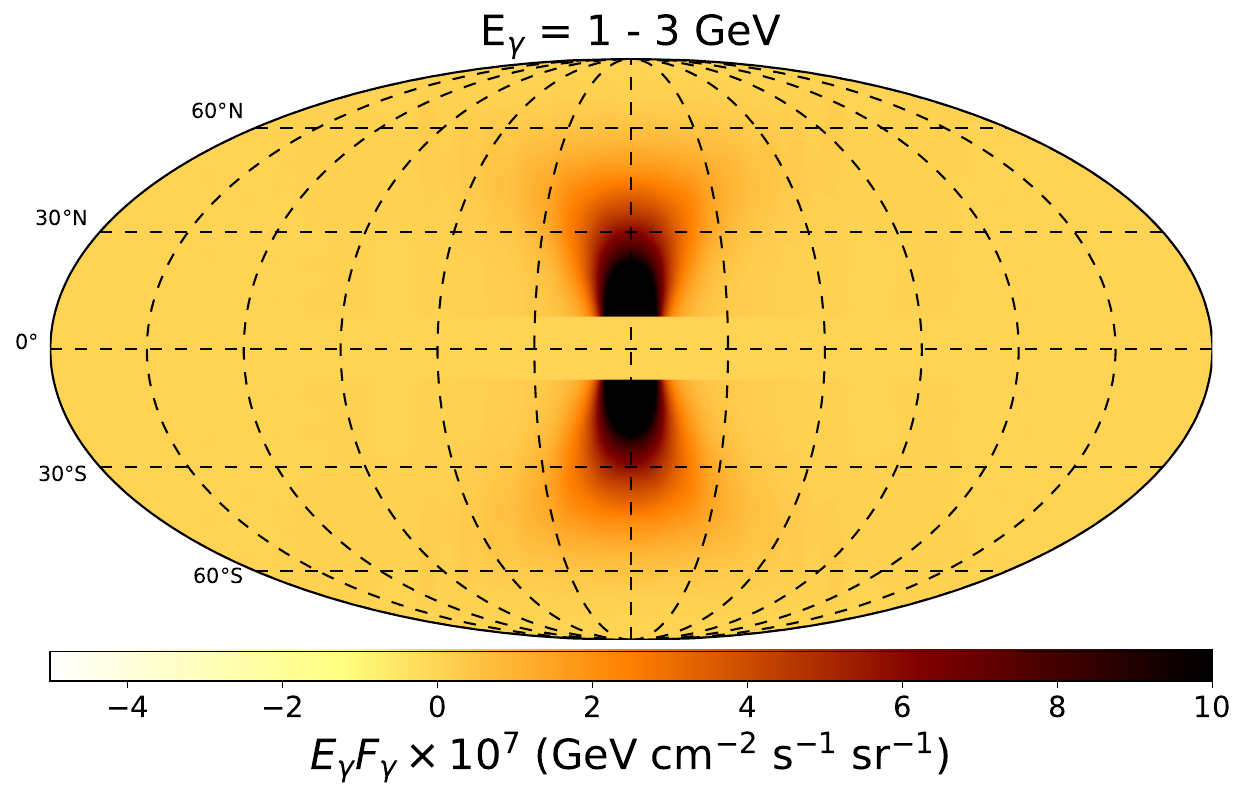}
    \caption{The 1-3~GeV $\gamma$-ray emission map produced through the energy losses of 10-30~GeV CR. The map uses a Mollweide projection. The colour bar is linear, and the Galactic plane is masked with $E_{\gamma}F_{\gamma} = 0$ at $|b| < 10^{\circ}$. Both colours and mask provide a setup similar to the observational paper \citep[Fig.~30]{Ackermann_2012}. The resulting emission produces a bubble shape, broadly consistent with observations (i.e. a height of $\sim 50^{\circ}$ and a width of $\sim 40^{\circ}$).} 
    \label{fig:skymap}
\end{figure*}

To understand the resultant CR density distribution, it is useful to look at the ratio of CR transport times shown in Fig.~\ref{fig:_b_diff}. While advection plays an important role for CR transport, it is not overwhelmingly dominant over diffusion. This is because a Galactic breeze model, by definition, implies a small velocity in comparison with the gas sound speed or jet speed. Due to the anisotropic diffusion coefficient, CRs diffuse mainly along the magnetic field lines. The combination of advection, anisotropic diffusion, and the collimation of the injected CRs produces an elongated CR profile along the $z$-direction. 

The CR density distribution obtained for the case of anisotropic diffusion can be compared with that obtained for the case of isotropic diffusion shown in appendix~\ref{Appendix_iso} (see Fig.~\ref{fig:Density_Iso}). As the CR diffusion is not influenced by the magnetic field direction for the isotropic diffusion case, the density distribution is more spherical in shape than for the case of anisotropic diffusion. Consequently, the CR contour lines show that the density distribution reaches a shorter distance along the $z$-axis and extends further along the $R$-axis for the isotropic diffusion case. For both anisotropic and isotropic diffusion cases considered here, CRs are injected with $\theta_{1/2} = 20^{\circ}$. 

Likewise, in contrast to the collimated CR injection considered here, in \citet{Tourmente_2023} the CR injection was not collimated. From \citet[Fig.~3]{Tourmente_2023}, one can see that in comparison to Fig.~\ref{fig:Density_Iso}, due to CRs being injected with no collimation, the CR density distribution is again more spherical, extending less far in the $z$-direction and extending further in the $R$-direction.

\subsection{$\gamma$-ray emission map}\label{Gamma_ray_emission_map}

\begin{figure*}
\center
\includegraphics[width=.74\linewidth]{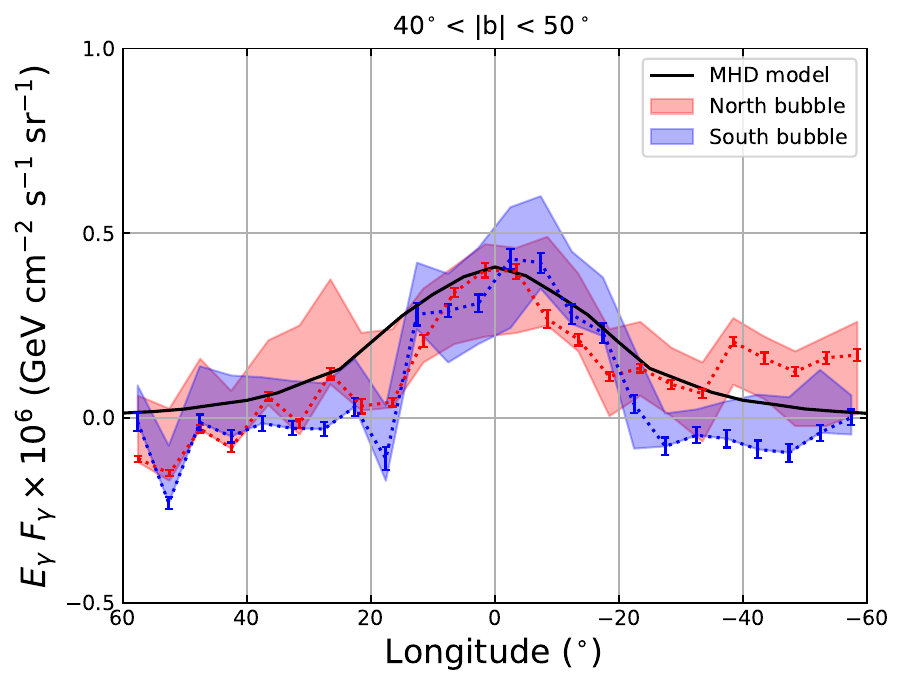}%
\caption{The energy flux distribution along the Galactic longitude for a Galactic latitude between $40^{\circ}$ and $50^{\circ}$. The solid black line represents the energy flux obtained through the CR transport simulation presented in this paper. The dashed red and blue lines, the error bar ranges and the shaded region have been provided by the Fermi-LAT observational results \citep{Fermi_2014}. The red colour represents the North Fermi bubble, and the blue colour represents the South Fermi bubble.}
\label{fig:40_50}
\end{figure*}

Using the obtained CR density spatial distribution, the subsequent $\gamma$-ray energy flux distribution can be determined (see Section \ref{Loss_time_scale}). Fig.~\ref{fig:skymap} presents the skymap of the $\gamma$-ray energy flux of 1-3~GeV photons produced through inelastic $pp$ collisions of 10-30~GeV CRs with the ambient Galactic halo gas. The skymap uses Galactic coordinates where the longitude, $l$, ranges from $0^{\circ}$ to $360^{\circ}$ and the latitude, $b$, ranges from $-90^{\circ}$ to $90^{\circ}$. The colour bar range has been chosen to be similar to the $\gamma$-ray emission maps presented in \citet{Fermi_2014}. For $|b| < 10^{\circ}$ a mask has been applied for which $E_{\gamma}F_{\gamma} = 0$, similar to Fig.~30 from \citet{Fermi_2014}. For $L_{\mathrm{CR}} = 1.3\times 10^{40}$~erg~s$^{-1}$ (with $D_{\perp} = 0.1 \times D_{\parallel}$ and $\theta_\mathrm{1/2}=20^{\circ}$), the Galactic breeze model is found to produce bilobal $\gamma$-ray emission whose brightness is comparable to that observed from the Fermi bubbles. 

Fig.~\ref{fig:skymap} can be understood by looking at Eq.~(\ref{subsec:Energy_loss}) and Eq.~(\ref{Energy_flux}). The $\gamma$-ray energy flux depends on $\tau_{\mathrm{loss}}$, i.e., on both the density distribution of the CRs and the Galactic halo gas. For the Galactic halo gas, the density distribution obtained from the MHD simulation has been used. As the propagation of the Galactic breeze does not disturb it (see Fig.~\ref{fig:Density}), this density distribution is largely consistent with expectation for isothermal gas in hydrostatic equilibrium. The $\gamma$-ray emission is proportional to both the CR density and the Galactic halo density. The $\gamma$-ray energy flux is then larger at the base of the bubbles than at the top, as the density distribution is larger at smaller Galactic radii. The bubbles reach $b\sim 55^{\circ}$. For the latitude range $40^{\circ} < b < 50^{\circ}$, $E_{\gamma}F_{\gamma} \approx 4\times 10^7$~GeV~cm$^{-2}$~s$^{-1}$~sr$^{-1}$ corresponding to the observations of the Fermi bubbles \citep{Fermi_2014}. At $b=10^{\circ}$, the simulation gives a bubble structure with a total width of $\sim 30^{\circ}$, consistent with the observations \citep{Carretti_2013, Fermi_2014}.

The analysis of the energy flux distribution is taken further in Fig.~\ref{fig:40_50}. This figure shows the $\gamma$-ray energy flux for $40^{\circ} < |b| < 50^{\circ}$ as a function of longitude for $-60^{\circ} < l < 60^{\circ}$. The solid black line represents the energy flux distribution obtained from our CR transport simulation. The dotted red and blue lines, error bars and red and blue shaded regions, for which the red corresponds to the North Fermi bubbles and the blue to the South Fermi bubbles, are obtained from \citep[Fig.~23]{Fermi_2014}. The lines and error bars shown in the figure correspond to the energy flux calculated using templates generated with the CR propagation code GALPROP. The shaded regions are computed for different configurations for the GALPROP template and local template proposed by the Fermi-LAT collaboration. In appendix \ref{App:30_40}, Fig.~\ref{fig:30_40} presents the energy flux distribution for $30^{\circ} < |b| < 40^{\circ}$. 

The CR injection luminosity in the simulations was set to a value of $L_{\mathrm{CR}} = 1.3\times 10^{40}$~erg~s$^{-1}$ in order for the corresponding value of the $\gamma$-ray brightness $E_{\gamma}F_{\gamma}$ to approximately match that of the observed Fermi bubbles (see the dotted red line in Fig.~\ref{fig:40_50}) at $l = 0^{\circ}$, $40^{\circ}<b<50^{\circ}$. For $E_{\gamma} = 1~-~3$~GeV, the total corresponding $\gamma$-ray luminosity produced by the bubbles is $L_{\gamma} = 2.7 \times 10^{37}$~erg~s$^{-1}$. The simulated $\gamma$-ray brightness is broadly consistent with the shaded red/blue regions, decreasing sharply between $0^{\circ}$ and $25^{\circ}$ of longitude, contrary to the hydrodynamic model, which produced a softer decrease \citep[see Fig.~5 from][]{Tourmente_2023}. Given the symmetry of our simulation setup, no north-south or east-west asymmetry is produced by our model. 

Once again, it is interesting to compare the results of the anisotropic diffusion case considered with those produced for the case of isotropic diffusion. In appendix~\ref{Appendix_iso}, Fig.~\ref{fig:iso_40_50} shows the $\gamma$-ray energy flux distribution for the case of isotropic diffusion. It is observed that in the case of isotropic diffusion, a more spherical shape for the $\gamma$-ray energy flux distribution is produced (see Fig.~\ref{fig:Map_Iso}), with the sharp edges of the Fermi bubbles not reproduced for the case of isotropic diffusion.

Furthermore, by comparing our results for the case of the collimated injection and isotropic diffusion (see appendix~\ref{Appendix_iso}) with the results for non-collimated injection and isotropic diffusion obtained in \citet{Tourmente_2023}, the effect of CR collimation on the resulting $\gamma$-ray energy flux distribution can be appreciated. From Fig.~\ref{fig:iso_40_50} and Fig.~5 from \citet{Tourmente_2023} one observes that the $\gamma$-ray energy flux for the case of collimated injection is more elongated in the $l$ direction and narrower in the $b$ direction, as expected.

\section{Discussion}\label{Discussion}

Our MHD and CR transport simulations allow us to calculate the fraction of the CR energy which subsequently goes on to power the Fermi bubble emission in $\gamma$-rays, a quantity known as the calorimetric fraction. For the CR luminosity used in this paper (see section \ref{Gamma_ray_emission_map}) a pure calorimetric exchange, i.e., 100\% of the CR luminosity is radiated as $\gamma$-rays, would correspond to $L_{\gamma} = \frac{1}{3}L_\mathrm{CR} = 4.3\times 10^{39}$~erg~s$^{-1}$, where the factor $1/3$ accounts for the $\gamma$-ray emission produced by $\pi^{0}$ \citep{Ackermann_2012, Wang_2018}. For our model, $L_{\gamma} = 2.7 \times 10^{37}$~erg~s$^{-1}$ (see section \ref{Gamma_ray_emission_map}), corresponding to a calorimetric fraction of $\sim 6\times 10^{-3}$. This implies that most of the CR energy is not radiated away in $\gamma$-rays from the Fermi bubble region, indicating then that these bubbles are non-calorimetric. However, it should be noted that the enhanced inner Galactic density, at least one order larger than the Galactic halo density at a Galactic radius of 300~pc, has not been included in our simulations. This larger density distribution in the central region would result in the production of further $\gamma$-ray emission from the central region, increasing the calorimetric fraction estimate.

For the model presented here, CRs have been injected into a solid cone at the Galactic centre region (see section \ref{CR_injection}). An injection of CRs into a hollow cone, however, is also motivated. Consideration of such a CR injection term is suggested from recent optical line emission observations of galactic outflows for several star-forming galaxies, which have shown that they are often limb-brightened line-emitting structures, potentially fed by rings of star formation \citep{Strickland_2000, Cecil_2001}. These results may indicate that galactic outflows possess hollow structures \citep{Veilleux_2005}. In such a scenario, CRs would propagate along a specific angular direction, forming a ring-shaped region, leading to anisotropies in the CR distribution. The centre of the conic structure will have a lower CR density than the limb, thus reducing the interactions in the central region. This implies less $\gamma$-ray emission from $pp$ collisions. Therefore, the edges would appear brighter than the central region. 

A further simplification used for our results is the assumption of a single temperature of the outflow gas. The outflow gas velocities obtained from our MHD simulations are somewhat smaller than the value of $\sim$300~km~s$^{-1}$ (at the critical radius), motivated by observations \citep{Lockman_2020}. For the hot Galactic halo gas, observations have indicated the presence of gas in an even hotter phase \citep{Das_2019a, Das_2019b, Das_2021, Gupta_2021, Rahul_2023}. A temperature profile, expressed through a multi-thermal phase model, may therefore be warranted to consider in the future. Additionally, further studies of the magnetic field distribution in the halo seem also warranted. The presence of a stronger azimuthal magnetic field than that considered here could lead to an increase in the outflow velocity at the critical radius.

\section{Conclusions and outlook}\label{Conclusion}

The Fermi bubbles have frequently been associated with a Galactic outflow. Following our previous work in which a hydrodynamic Galactic breeze (subsonic) model was employed to describe the Fermi bubbles \citep{Tourmente_2023}, we have here extended this description using MHD simulations. For our fiducial model, an initial radial magnetic field at the inner radial boundary of 20~$\mu$G and an azimuthal magnetic field of 2~$\mu$G are adopted. This MHD simulation setup leads to a velocity profile largely similar to that obtained previously for our hydrodynamic model. Both the velocity and magnetic field profiles from our MHD simulation of the Galactic halo have subsequently been used as inputs for our CR propagation simulation. Our transport of CRs in the Galactic halo, via both advection and anisotropic diffusion, gave rise to $\gamma$-ray emission through inelastic CR collisions with the low-density hot Galactic halo gas.

The distribution of $\gamma$-ray emission produced from our model is plotted and compared against observational results by Fermi-LAT of the Fermi bubble regions. Additionally, our $\gamma$-ray emission result is compared with the results obtained for both the case of isotropic diffusion and for our previous hydrodynamic model \citep{Tourmente_2023}. These comparisons have highlighted three main aspects about our results. First, anisotropic CR diffusion can play a key role in shaping the spatial distribution of the $\gamma$-ray emission subsequently produced through CR energy losses. Due to the anisotropy of the diffusion coefficient, CRs follow relatively tightly the magnetic field lines that are themselves tied to the outflow. The effect of this, seen from a comparison of Figs.~\ref{fig:skymap} \& \ref{fig:40_50} with \citet[Figs.~4\&5]{Tourmente_2023}, is to produce synthetic $\gamma$-ray emission maps which are more closely in agreement with the observations from the Fermi satellite. 
Secondly, the inclusion of a half opening angle of $20^{\circ}$ for the injection of CRs leads to the CR distribution taking a more pronounced conical shape.

Finally, the CR luminosity needed to match the simulated $\gamma$-ray emission brightness with that of observations and the resulting calorimetric fraction of the CRs powering the Fermi bubbles $\gamma$-ray emission can be deduced. We find $L_\mathrm{CR} = 1.3\times 10^{40}$~erg~s$^{-1}$ and $L_{\gamma} = 2.7\times 10^{37}$~erg~s$^{-1}$. The inferred calorimetric fraction is therefore $\sim 6\times 10^{-3}$. This implies that the Fermi bubbles are far from calorimetric, with most of the CR energy escaping rather than being converted into radiation. 

In this paper we have proposed a model that offers promising results and motivates further investigations. Interesting avenues of further research include a broader comparison of the model's predicted $\gamma$-ray energy spectrum, as well as an investigation into the bubble emission out within the extended halo region within the Galactic virial radius \citep{Taylor:2014hya,Gabici:2021rhl}. Additionally, similar comparisons of the expected bubble emission around other nearby galaxies appear warranted and timely \citep{Pshirkov:2016qhu,Yang_2024}.

\section*{Acknowledgements}
The authors would like to thank the anonymous reviewer
for their valuable comments which helped to improve the
manuscript. OT and AMT acknowledge support from DESY (Zeuthen, Germany), a member of the Helmholtz Association HGF. DRL would like to acknowledge that this publication has emanated from research conducted with the financial support of Taighde {\'E}ireann – Research Ireland under Grant number 21/PATH-S/9339.

\section*{Data Availability}
The data underlying this article will be shared on reasonable request to the corresponding author.

\bibliographystyle{mnras}
\bibliography{sample.bib}

%\printbibliography

\appendix 

\section{Effects of a magnetic field distribution on a thermally driven outflow}\label{App:Bfield}

An introduction to the topic of a hydrodynamical spherically symmetric outflow can be found in chapter three of \citet{lamers_1999} based on the work of \citet{Parker_1958, Parker_1965}. We focus only here on a description for an MHD case. 

By considering a steady-state model, the expression for the gradient density from Eq.~(\ref{Mass_conservation}) can be included in the Eq.~(\ref{Momentum_conservation}). For a purely radial symmetry and after developing this last equation, the following expression for the radial component is

\begin{equation*}
    v_r\frac{dv_r}{dr} - \frac{c^2_s}{v_r}\frac{dv_r}{dr} - \frac{2c^2_s}{r} + \frac{d\Phi_{\mathrm{tot}}}{dr} - \frac{v^2_{\phi}}{r} + \frac{B_{\phi}}{4\pi \rho r}\frac{d}{dr}\left(r B_{\phi}\right) = 0,
    \label{N_MHD}
\end{equation*}
where $r$ is the spherical radius, $v_r$ is the radial velocity, $v_{\phi}$ is the azimuthal velocity, and $c_s$ is the thermal velocity, which is constant because an isothermal model is considered. $B_{\phi}$ is the azimuthal component of the magnetic field. This expression can be written into a more convenient form to get an intuition of the outflow evolution,

\begin{equation}
\frac{1}{v_r}\frac{dv_r}{dr} = \frac{1}{r}\left(\frac{2c^2_s - r\frac{d\Phi_{\mathrm{tot}}}{dr} + v^2_{\phi} - \frac{B_{\phi}}{4\pi \rho}\frac{d}{dr}\left(r B_{\phi}\right)}{v^2_r - c^2_s}\right).
\label{MHD_case}
\end{equation}
Assuming a hydrodynamic model, with $B_{\phi}=0$ and $v_{\phi}=0$, the numerator on the right-hand side of this expression goes to zero when $r\frac{d\Phi_{\mathrm{tot}}}{dr} = 2c^2_s$. The radius where such an equality occurs is called the critical radius, $r_{\rm c}$. Close to the Galactic centre the outflow is gravitationally bound. A subsonic outflow will then increase until it reaches the critical radius, after which it continuously decelerates. The addition of a rotational velocity or a magnetic field can affect the position of $r_{\rm c}$. The rotational velocity has the effect of reducing the gravitational pressure and thus shifting the critical radius to the Galactic centre. The azimuthal magnetic field can have different effects depending on its distribution. 

Considering a dependency with $r$ only, $B_{\phi}$(r) can be written as
\begin{equation*}\label{B_phi}
    B_{\phi}(r) = B_{\phi_{s}}\left(\frac{r_s}{r}\right)^{\beta},
\end{equation*}
where $r_s$ is the scale radius, $B_{\phi_{s}}$ is the strength of the azimuthal magnetic field at $r_s$ and $\beta$ represents the power-law index. Looking at Eq.~(\ref{MHD_case}), the position of $r_{\rm c}$ will change depending on the power-law slope of the $B_{\phi}$-field component, 
\begin{equation*}
r_c = \frac{2c^2_s + v_{\phi}^2 - \frac{B^2_{\phi}}{4\pi \rho}\left(1-\beta\right)}{\frac{d\Phi_{\mathrm{tot}}}{dr}} \rightarrow
\begin{cases}
\beta = 1 & \Rightarrow B^2_{\phi}\left(1-\beta\right) = 0\\
\beta > 1 & \Rightarrow B^2_{\phi}\left(1-\beta\right) < 0\\
\beta < 1 & \Rightarrow B^2_{\phi}\left(1-\beta\right) > 0.
\end{cases}
\end{equation*}
For $\beta = 1$, the magnetic pressure and the magnetic tension cancel each other, and the position of the critical radius is not modified by the presence of the magnetic field. For $\beta > 1$, the magnetic pressure is more important than the magnetic tension, and $r_{\rm c}$ is moved closer to the Galactic centre. For $\beta < 1$, the magnetic pressure is less important than magnetic tension, and $r_{\rm c}$ moves farther from the Galactic centre. This last case allows for increasing the thermal pressure for keeping a critical radius at 1~kpc. This competition between the influence of the magnetic pressure and the magnetic tension has been previously explored to deduce its influence on the rotation of the Galactic disc \citep{Sanchez_2013, Elstner_2014}.

\section{Analytic expression for $\mathbf{B_{\phi}(r)}$}\label{App:Bphi}

Assuming a steady-state condition for the induction equation, Eq.~(\ref{Faraday_law}), gives,

\begin{equation*}
    \nabla \times \left(\textbf{\textit{v}}\times \textbf{B}\right) = 0.
\end{equation*}
This equation can be used to obtain an expression for $B_{\phi}(r)$ that depends on $v_r(r)$, $v_{\phi}(r)$, and $B_r(r)$,

\begin{equation*}
    \frac{1}{r}\frac{d}{dr}\left(r\left(v_rB_{\phi} - v_{\phi}B_r\right)\right) = 0,
\end{equation*}
leading to

\begin{equation*}
    r\left(v_rB_{\phi} - v_{\phi}B_r\right) = r_0\left(v_{r_{0}}B_{\phi_{0}} - v_{\phi_{0}}B_{r_{0}}\right).
\end{equation*}
To have a better understanding of the relationship between the different components, several power-law distributions can be considered

\begin{align*}
v_r &= v_{r_{0}}\left(\frac{r_0}{r}\right)^{\alpha}, \\[1ex]
v_{\phi} &= v_{\phi_{0}}\left(\frac{r_0}{r}\right), \\[1ex]
B_{r} &= B_{r_{0}}\left(\frac{r_0}{r}\right)^2,
\end{align*}

giving

\begin{equation*}
B_{\phi} = \left[B_{\phi_{0}} + B_{r_{0}}\frac{v_{\phi_{0}}}{v_{r_{0}}}\left(\frac{r^2_0}{r^2} - 1\right)\right]\left(\frac{r_0}{r}\right)^{1-\alpha}.
\end{equation*}
From this expression, one can see that it is not necessary to have $B_{\phi_{0}} \neq 0$ for $B_{\phi}(r)$ to exist. If $B_{\phi_{0}} = 0$, $v_{\phi_{0}}$ and $B_{r_{0}}$ must be different from zero. Moreover, the combination $B_{r_{0}}v_{\phi_{0}}$ decreases $\propto r^{-2-1+\alpha}$. A large value $B_{r_{0}}$ or $v_{\phi_{0}}$ is then necessary for $B_{\phi}(r)$ to reach a few $\mu$G $1~\mathrm{kpc} \leq z \leq 10~\mathrm{kpc}$. Also, a too large value for $v_{\phi_{0}}$ would disrupt the propagation of the outflow, as it would compete with $v_{r_{0}}$. It must also be noted that the distribution of $B_{\phi}(r)$ depends mostly on the index $\alpha$ representing the power-law distribution for $v_r(r)$. For the Galactic breeze profile (see Fig.~\ref{fig:_c}) for $z > 1$~kpc, $\alpha \approx 1$. The distribution of $B_{\phi}(r)$ is then not steep and looks quasi-constant along $r$. 

\section{Effect of compression and advection terms for the induction equation}\label{App_Induction}

\begin{figure}
\center
\includegraphics[width=1.\linewidth]{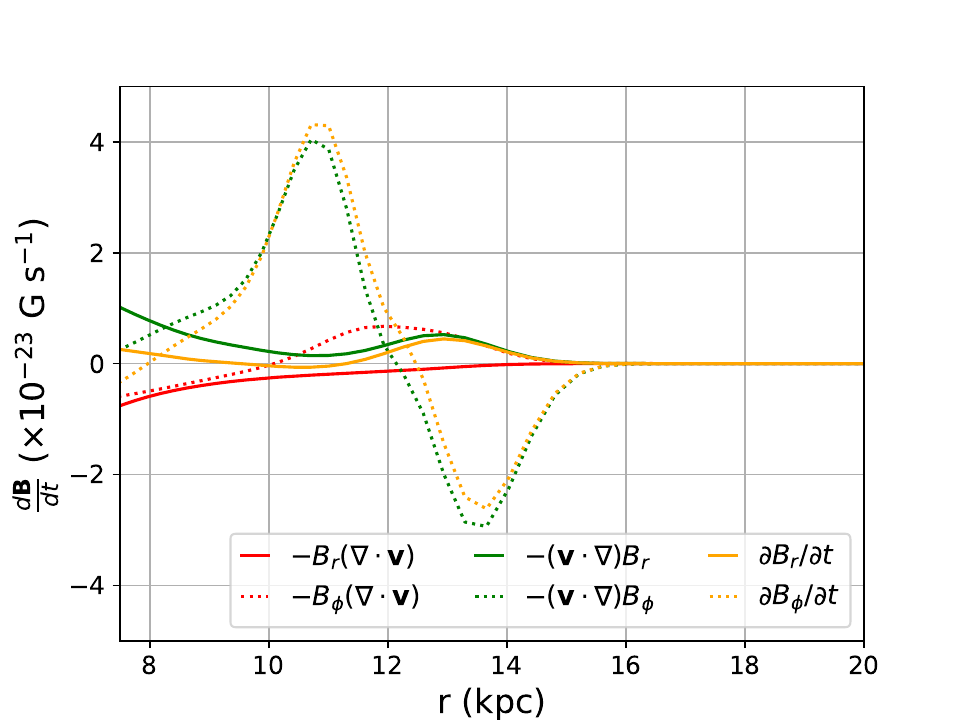}%
\caption{Distribution of the induction equation components (see Eq.~\ref{induction_components}) for the last time step, corresponding to compression (in red) and advection (in green) and the total of those two terms (in orange) along $r$. Solid lines correspond to the radial term and dotted lines correspond to the azimuthal term.}
\label{fig:Induction}
\end{figure}

The induction equation (Eq. \ref{Faraday_law}) can be re-expressed as
\begin{equation}
    \frac{\partial \bfm{B}}{\partial t} = -\bfm{B}\left(\nabla \cdot \bfm{v}\right) - \left(\bfm{v} \cdot \nabla\right)\bfm{B} + \left(\bfm{B} \cdot \nabla\right)\bfm{v}\label{induction_components},
\end{equation}
where the terms on the right-hand side correspond to compression, advection, and stretching of the magnetic field, respectively. Fig. \ref{fig:Induction} presents the distribution of the different terms corresponding to the compression (in red), the advection (in green) and the total of those terms (in orange). The solid lines correspond to the radial term and dotted lines correspond to the azimuthal term, respectively. The stretching term has been omitted as it is negligible in comparison of the two other terms. The shell is predominantly formed by the azimuthal advection term. It should be noted that the time scale on which $B_{\phi}$ changes, that is $\sim $~Gyr scale, is significantly larger than the dynamical age of the Fermi bubbles, 300 Myr for our simulations, ensuring that the CR distribution reaches a steady-state out of a distance of 10~kpc. 

\section{The expression of the diffusion coefficient}\label{Appendix_B}

A 2D cylindrical symmetry ($R$, $z$) is considered here. Assuming that the parallel diffusion is done along the $z$-axis, $D_{\parallel}=D_{zz}$ and then the perpendicular is done along the $R$-axis, $D_{\perp} = D_{RR}$. The cross-diffusion terms are considered to be zero, $D_{Rz} = D_{zR} = 0$. 

The transformation of a tensor is expressed by 
\begin{equation*}
T = R^{-1}T'R,
\end{equation*}
where $T$ and $T'$ represent a tensor and $R$ is the transformation matrix, here
\begin{equation*}
R = 
\begin{pmatrix}
\cos{\Theta} & \sin{\Theta}\\
-\sin{\Theta} & \cos{\Theta}
\end{pmatrix}.
\end{equation*}
The transformation of the diffusion tensor leads then to the following expression
\begin{equation*}
\begin{pmatrix}
D_{RR} & D_{Rz}\\
D_{zR} & D_{zz}
\end{pmatrix}
= 
\end{equation*}
\begin{equation*}
\begin{pmatrix}
\cos{\Theta} & -\sin{\Theta}\\
\sin{\Theta} & \cos{\Theta}
\end{pmatrix}
\begin{pmatrix}
D_{\perp} & 0\\
0 & D_{\parallel}
\end{pmatrix}
\begin{pmatrix}
\cos{\Theta} & \sin{\Theta}\\
-\sin{\Theta} & \cos{\Theta}
\end{pmatrix},
\end{equation*}
giving
\begin{align*}
D_{RR} &= D_{\perp}\cos^2 {\Theta} + D_{\parallel}\sin^2{ \Theta} = D_{\perp}\left(1 - \sin^2 {\Theta}\right) + D_{\parallel}\sin^2 {\Theta}\\
D_{Rz} &= D_{zR} = \left(D_{\parallel} - D_{\perp}\right)\cos{\Theta} \sin{\Theta}\\
D_{zz} &= D_{\perp}sin^2{\Theta} + D_{\parallel}\cos^2{\Theta} = D_{\perp}\left(1 - \cos^2{\Theta}\right) + D_{\parallel}\cos^2{\Theta}.
\end{align*}
The functions $\cos{\Theta}$ and $\sin{\Theta}$ can be related to the components of the total magnetic field as
\begin{align*}
\sin{\Theta} &= \frac{B_{R}}{|\textbf{B}|}  \\
\cos{\Theta} &=\frac{B_{z}}{|\textbf{B}|}.
\end{align*}
Incorporating this into the above leads to the general diffusion tensor expression,
\begin{equation*}
\bfm{D} = D_{\perp}\left(\bfm{I} - \frac{\bfm{B}}{|\bfm{B}|}\frac{\bfm{B}}{|\bfm{B}|}\right) + D_{\parallel}\frac{\bfm{B}}{|\bfm{B}|}\frac{\bfm{B}}{|\bfm{B}|}.
\end{equation*}

\section{Energy flux distribution for latitude between 30$^{\circ}$ and 40$^{\circ}$}\label{App:30_40}

\begin{figure}
\center
\includegraphics[width=1.\linewidth]{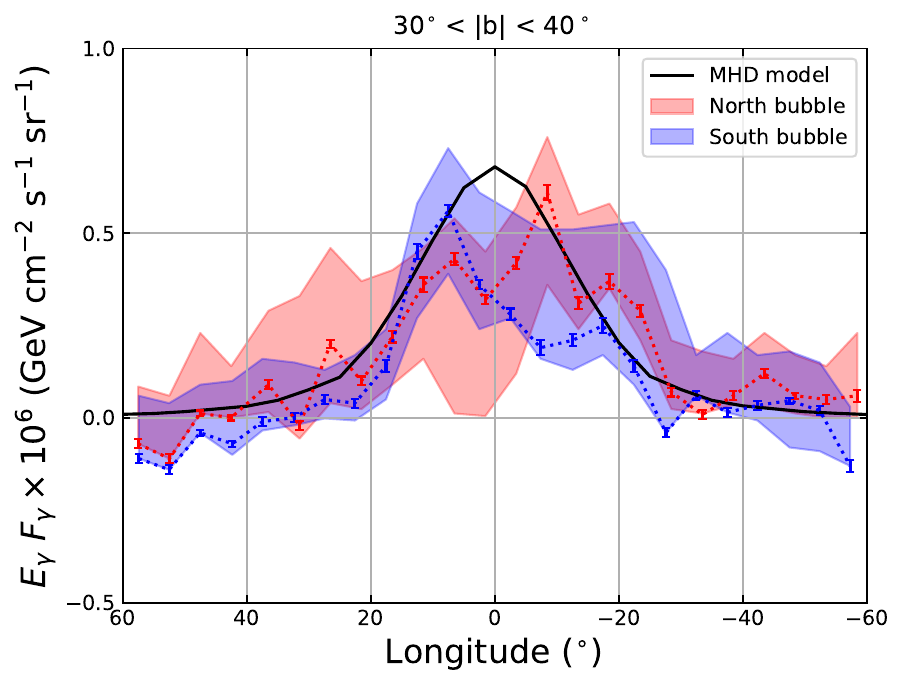}%
\caption{Energy flux distribution along the Galactic longitude for a Galactic latitude between $30^{\circ}$ and $40^{\circ}$. The line styles, shaded regions and normalisation are identical to Fig.~\ref{fig:40_50}.}
\label{fig:30_40}
\end{figure}

In Fig.~\ref{fig:30_40} a direct comparison is made between the simulation result and the observational data provided by the Fermi-LAT collaboration \citep{Fermi_2014} for $30^{\circ} \leq b \leq 40^{\circ}$. The line styles and shaded regions are similar to the case presented in Section \ref{Gamma_ray_emission_map} for $40^{\circ} \leq b \leq 50^{\circ}$. The black line, i.e., the energy flux, $E_{\gamma}F_{\gamma}$ has been normalised to match with the red line at a longitude of 0$^{\circ}$ for a latitude $40^{\circ} \leq b \leq 50^{\circ}$ (see Fig.~\ref{fig:40_50}). This normalisation corresponds to $L_{\mathrm{CR}} = 1.3\times 10^{40}$~erg~s$^{-1}$ for a gamma-ray energy range between 1 and 3~GeV. The brightness is larger than the upper limit of both the shady red and blue regions at a longitude of $0^{\circ}$. %The observations show a bending of the bubbles that is not reproduced by the presented model. 

\section{Gamma-ray energy flux distribution for an isotropic diffusion model}\label{Appendix_iso}

Similar to the Section \ref{Results}, treating the results for the anisotropic diffusion model, this appendix covers the isotropic diffusion model. The setups for both MHD simulations and CR transport are identical to the ones presented in Section \ref{sec:Galactic_outflow} and Section \ref{sec:CR_transport}. The only difference is the isotropy of the diffusion that is obtained by considering the equality $D_{\parallel} = D_{\perp}$, where $D_{\parallel}$ is defined by Eq. (\ref{Dc_inhomo}). 

\begin{figure}
    \center
    \includegraphics[width=1.\linewidth]{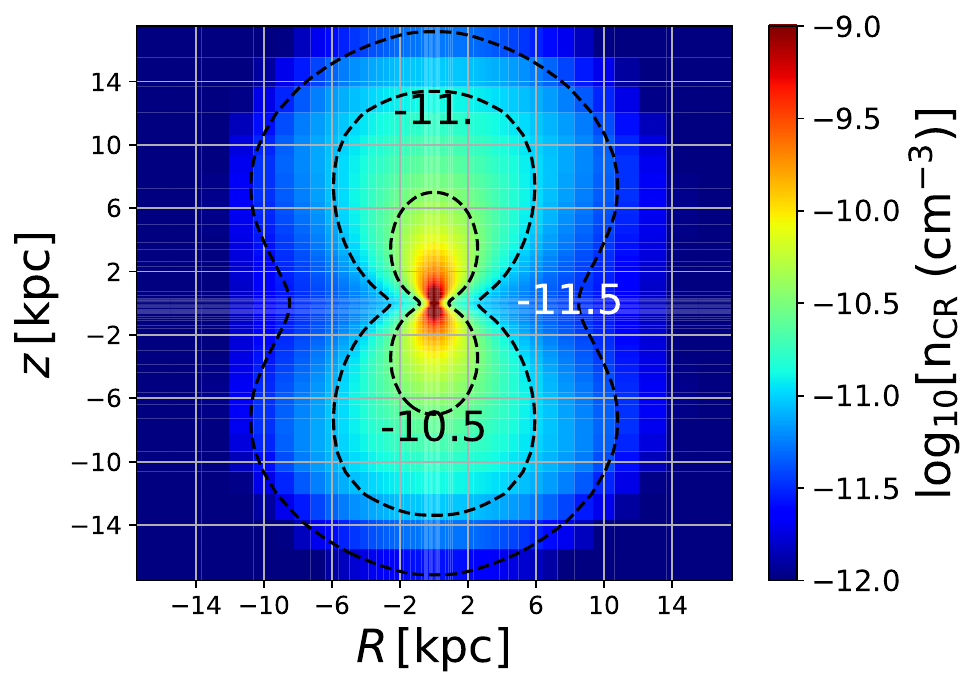}
    \caption{Spatial CR density distribution in the Galactic halo for an energy range of $E_{\mathrm{CR}} = 10~-~30$~GeV for the isotropic diffusion model. The colours and contour lines are the same as for Fig.~\ref{fig:Density_distrib}.}
    \label{fig:Density_Iso}
\end{figure}

The CR density distribution can be seen on Fig.~\ref{fig:Density_Iso}. The colour bar range and line styles are identical to Fig.~\ref{fig:Density_distrib}. By comparing Fig.~\ref{fig:Density_Iso} with Fig.~\ref{fig:Density_distrib} one can see that the isotropic model leads to a different distribution. As CRs are not constrained to move along magnetic lines, they diffuse more broadly. This effect can be clearly seen by looking at the farthest contour line from the map centre for which $n_{\mathrm{CR}} = 3\times 10^{-11}$~cm$^{-3}$. At a height of 0~kpc, the contour lines reach a half width of $\sim$9~kpc when for the anisotropic diffusion model, this contour line reaches a width of $\sim$3~kpc. This broader diffusion also has the consequence of reducing the height reached by the contour line. This effect can be seen by looking at the contour line the closest from the centre map, for which $n_{\mathrm{CR}} = 3\times 10^{-10}$~cm$^{-3}$. For $R$ = 0~kpc, this contour line reaches a height of $\sim$7~kpc where it reaches almost 15~kpc for the anisotropic diffusion model.  

\begin{figure}
    \center
    \includegraphics[width=1.\linewidth]{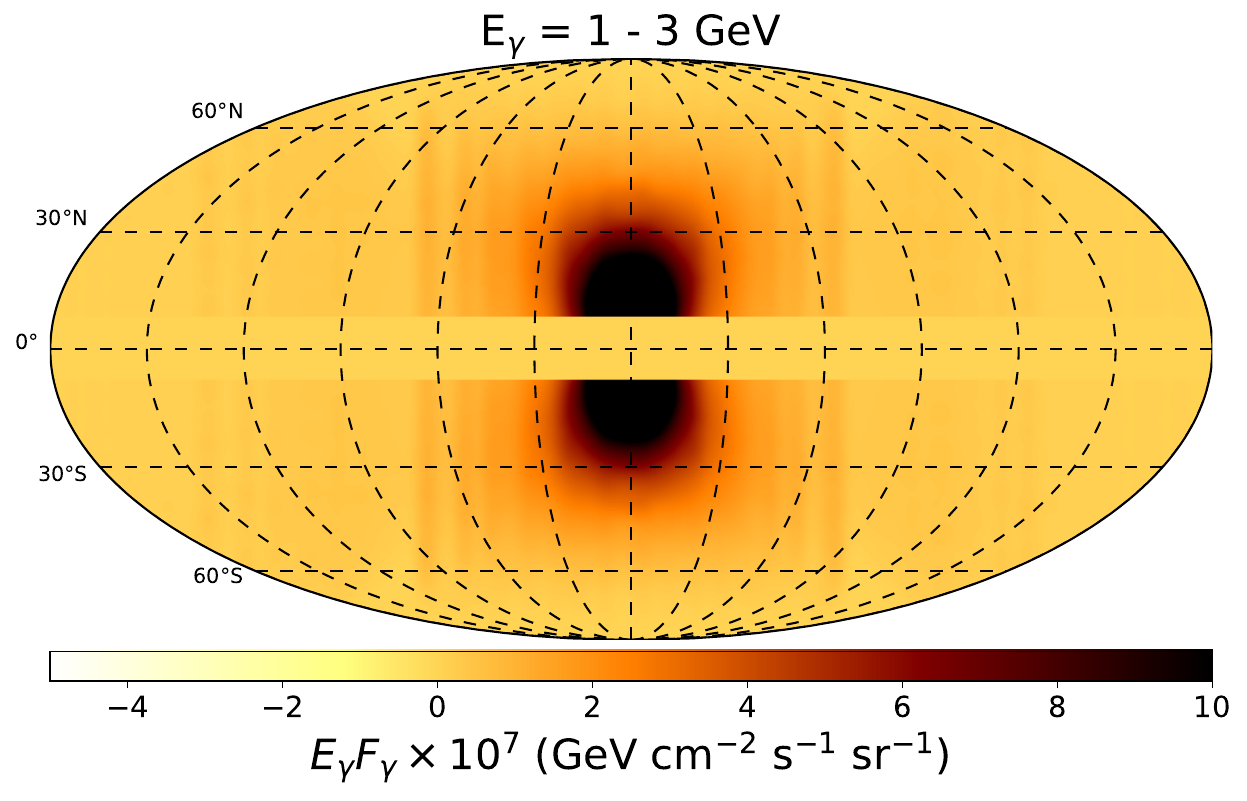}
    \caption{$\gamma$-ray emission map for an energy range of 1 to 3~GeV obtained from an isotropic diffusion of CRs. A Mollweide projection has been used. The colour bar and mask are similar to Fig.~\ref{fig:skymap}. The $\gamma$-ray emission has been produced by $pp$ collisions following the advection of CRs into the hydrostatic density distribution of the Galactic halo. The resulting emission produces a bubble shape for one half hemisphere that is larger than for an anisotropic CR diffusion (see Fig.~\ref{fig:skymap}).}
    \label{fig:Map_Iso}
\end{figure}

\begin{figure}
\center
\includegraphics[width=1.\linewidth]{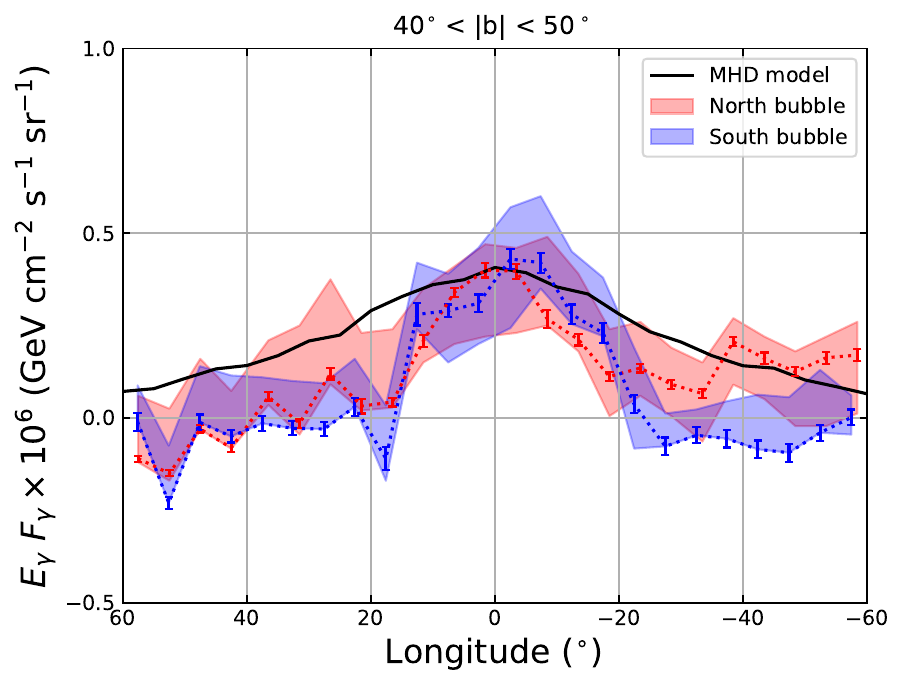}%
\caption{$\gamma$-ray energy flux distribution as a function of Galactic longitude, $l$, for a Galactic latitude between $40^{\circ}$ and $50^{\circ}$, assuming isotropic CR diffusion. The line styles and shaded regions are the same as for Figs.~\ref{fig:40_50} and \ref{fig:30_40}.}
\label{fig:iso_40_50}
\end{figure}

From the CR density distribution, the skymap of the $\gamma$-ray energy flux is obtained. This skymap can be seen with Fig.~\ref{fig:Map_Iso}. The choice of coordinates and colour bar is identical to Fig.~\ref{fig:skymap}.  Following the CR density distribution, the energy flux distribution shows a more spherical shape and broader distribution than the anisotropic diffusion model. Fig.~\ref{fig:iso_40_50} presents the $\gamma$-ray distribution for a latitude between $40^{\circ}$ and $50^{\circ}$. The lines and dashed regions are similar to Fig.~\ref{fig:40_50}. The broader distribution of CRs leads to a less steep distribution of the energy flux along the longitude. Consequently, it does not match as well as the anisotropic diffusion model. The energy flux distribution has been normalised to match with the red line at a longitude of $0^{\circ}$. This normalisation corresponds to a CR luminosity of $\sim 3 \times 10^{40}$~erg~s$^{-1}$. The total gamma-ray luminosity for the gamma-ray energy range $E_{\gamma} = 1~-~3$~GeV, is $L_{\gamma} = 5.4\times 10^{37}$~erg~s$^{-1}$.

\end{document}